# REDUCING THE POSSIBILITY OF SUBJECTIVE ERROR IN THE DETERMINATION OF THE STRUCTURE-FUNCTION-BASED EFFECTIVE THERMAL CONDUCTIVITY OF BOARDS

*Ernő Kollár, Vladimír Székely*

BUTE, Department of Electron Devices, H-1521 Goldmann Gy. tér 3, Budapest, Hungary
<kollar|szekely@eet.bme.hu>

**ABSTRACT**

The thermal response function given to a unit-step dissipation accurately characterizes the thermal system. Instead of the thermal response function the so-called structure function describing three-dimensional as the equivalent model of one-dimensional heat-spreading, created from the thermal response function with the help of complex mathematical procedures, is often used. Using the structure function the partial thermal capacity and partial heat resistance of certain elements of the thermal system can be identified.

If the geometrical measurements of a thermal system of simple geometry and homogeneous material (such as a homogeneous rod or board, etc.) are known, the coefficient of thermal conductivity of the material in question can be determined from two points of the structure function at 2-5 per cent of accuracy.

In this paper a method is presented which applies a wide range/section instead of two points of the cumulative structure function to determine the thermal coefficient, thus reducing the subjective error deriving from the selection of the two points.

The above method is presented and illustrated in simulated as well as measured thermal transient responses.

## 1. INTRODUCTION

The structure functions are graphical representations of the RC-model (Cauer-network) of the thermal system (Figure 1).

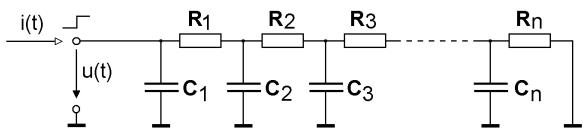

Figure 1: The RC-model of the thermal system.

The thermal resistance between the *n*-th element of the model network and the heat source (driving point) is:

$$R_\Sigma(n) = \sum_{i=1}^{n} R_i, \qquad (1)$$

and the cumulative thermal capacitance is:

$$C_\Sigma(n) = \sum_{i=1}^{n} C_i, \qquad (2)$$

where $R_i$ and $C_i$ denote the element values of the *i*-th stage of the Cauer-type model network. Plotting the $C_\Sigma$ vs. $R_\Sigma$ values results in the cumulative structure function (Figure 2).

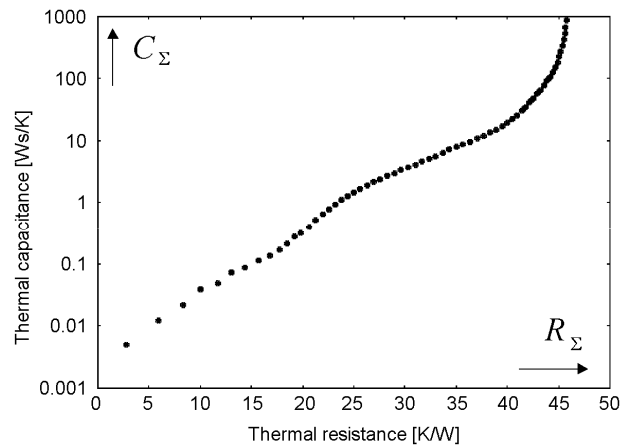

Figure 2: The cumulative structure function.

The cumulative structure function characteristic of board-like materials of homogeneous isotropic conductivity (Figure 3).

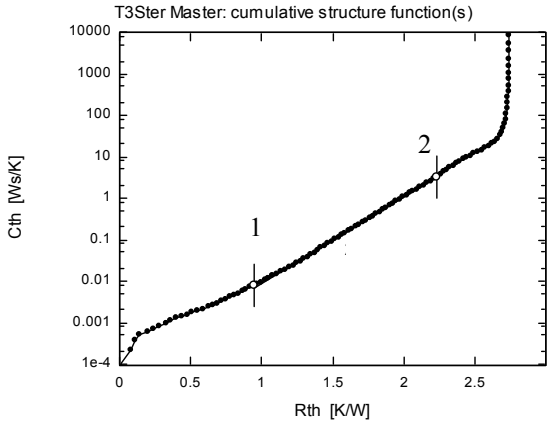

Figure 3: The cumulative structure function of homogeneous board.

In order to determine the conductivity value on the linear range from the integrated structure function [1]:

$$w\lambda = \frac{1}{4\pi} \cdot \frac{\ln\left(\frac{C_{\Sigma 2}}{C_{\Sigma 1}}\right)}{R_{\Sigma 2} - R_{\Sigma 1}}, \quad (3)$$

where $w$ is the thickness of the board and the $R_\Sigma$ and $C_\Sigma$ indexes indicate the values belonging to the selected points.

The main aspect of the selection of the two points is that both should be within the middle, linear (lambda measuring) range and should not be identical.

Our simulations carried out by the SunRed [2] simulation program showed that considering the criteria of the above mentioned method as a basis and keeping in mind the aspects of accuracy these criteria are not enough. In the selection of the two points there always remains a certain degree of subjectivity, the subjectivity of the person carrying out the evaluation, which appears as an error when determining the thermal conductivity.

Therefore we have elaborated a method to reduce subjective error, which applies a wide range/section instead of two points of the cumulative structure function to determine the thermal coefficient [3].

It is an additional advantage of the method that it determines the inflection point of the selected range more accurately, than determining it from the second derivative according to the $R_{th}$.

## 2. THE λ-FINDER ALGORITHM

The λ-finder algorithm determines the thermal conductivity coefficient from the integrated structure function of the board-like materials. The algorithm examines how the steepness of the graph (curve) changes if the symmetrical immediate neighborhood of the given point in the integrated structure function is symmetrically extended.

The steps of the algorithm are the following:
- Conductivity values are determined in each point of the range according to (4)

$$\lambda_n(R_{th}) = \frac{1}{w} \cdot \frac{\pi}{4} \cdot \ln\left(\frac{\frac{C_{th(+n)}}{C_{th(-n)}}}{R_{th(+n)} - R_{th(-n)}}\right), \quad (4)$$

where $w$ is the thickness of the board, and n in the indexes of $R_{th}$ and $C_{th}$ indicates the $R_{th}$ and $C_{th}$ values plus ($+n$) samples towards the higher values and minus ($-n$) samples towards the smaller values from the point examined. ($n = 1\ldots20$).

- From the conductivity values belonging to a certain point of the range $\lambda_m(R_{th})$ average(s) (5) and $\delta_m(R_{th})$ relative deviation(s) (6) are determined.

$$\overline{\lambda}_m(R_{th}) = \frac{1}{m}\sum_{n=1}^{m}\lambda_n(R_{th}), \quad (5)$$

$$\delta_m(R_{th}) = \frac{1}{\overline{\lambda}_m(R_{th})}\sqrt{\frac{\sum_{n=1}^{m}(\lambda_n(R_{th}) - \overline{\lambda}_m(R_{th}))^2}{(m-1)}} \quad (6)$$

- Describe the $\delta_m(R_{th})$ relative deviations and $\lambda_m(R_{th})$ average thermal conductivity(ies) in relation to $R_{th}$ thermal resistance.
- We seek minimal values in the $\delta_m(R_{th})$ deviation functions.
- Where the $\delta_m(R_{th})$ functions together give the smallest minimum, this $\delta_m(R_{th})$ value belonging to the $R_{th}$ point gives the effective thermal conductivity on $\lambda_m(R_{th})$ function as well as the inflection point of the integrated structure function.

Note: The $n_{max} = 20$ and $m$ values are experimental ones. The algorithm does not exclude the use of any other $n_{max}$ and $m$ values.

## 3. EXAMPLE I.

In this example we started from the simulation of the thermal transient heat response of the 1-mm-thick homogeneous copper board in vacuum. We created an integrated structure function from the gained thermal response function using the *T3Ster-Master* [4] program.

The program determines conductivity from two points of the integrated structure function selected at random and a board thickness given at random. The calculated thermal conductivity of the sample may differ from the 390 W/mK value set as simulation parameter even with 1.5 per cent, depending on the two selected points (Figure 4).

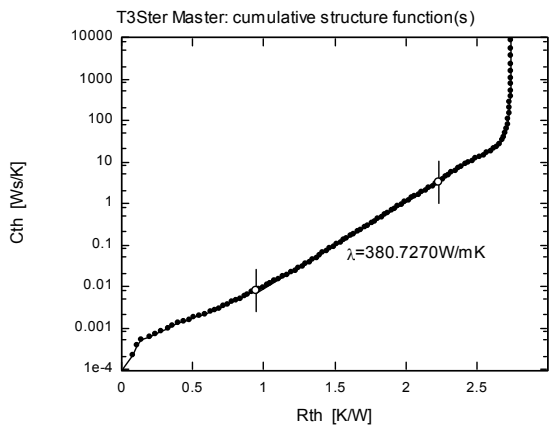

Figure 4: The integrated structure function of the 1-mm-thick ECU57 sample 'simulated in vacuum'

Let us employ the λ-finder algorithm for the calculated integrated structure function and describe the $δ_m(R_{th})$ relative deviations in relation to the $R_{th}$ thermal resistance.

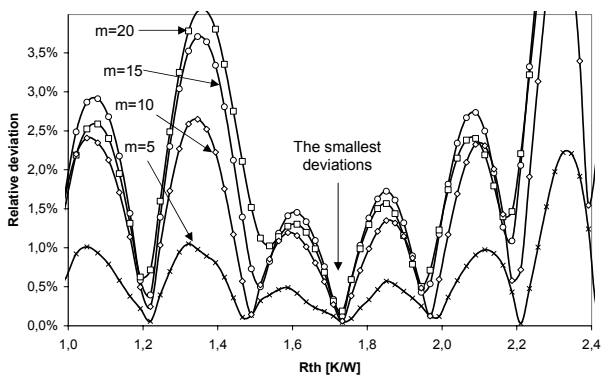

Figure 5: The $δ_m(R_{th})$ relative deviation functions of the sample

The smallest deviations are gained at $R_{th}$ = 1.73 W/mK. Describing the $λ_m(R_{th})$ average thermal conductivity values in relation to $R_{th}$ heat resistance we can read the values belonging to the above thermal conductivity resistance. Figure 6 shows the function and Chart 1 summarizes the values read.

The smallest minimums of the relative deviation functions designate accurately that $R_{th}$ point where the determination of the thermal conductivity value is advisable. With the help of the smallest minimum points found in the relative deviation function of thermal conductivity we are able to prevent the appearance of subjective error at the determination of thermal conductivity value derived from two points.

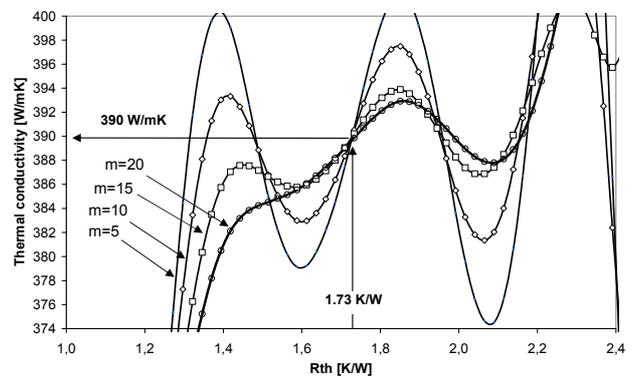

Figure 6: The $λ_m(R_{th})$ average thermal conductivity functions of the sample

In certain cases it might be necessary to determine the inflection point of the cumulative structure function. How-ever, the determination from the second derivative according to the $R_{th}$ is not always a preferable solution. The double derivation may increase the 'noise' the function so much that we can select the inflection point only in a subjective manner.

The λ-finder algorithm smooths down this 'noise' and provides the inflection point of the cumulative structure function at the smallest minimum point of the deviation functions.

| M | 5 | 10 | 15 | 20 |
|---|---|---|---|---|
| $δ_m$ [%] | ~~0.04~~ | 0.07 | 0.13 | 0.20 |
| $λ_m$ [W/mK] | ~~403.96~~ | 390.5 | 390.2 | 389.9 |
| $R_{th}$ [K/W] | ~~2.21~~ | 1.73 | 1.73 | 1.73 |

Chart 1: Thermal conductivity of the ECU-57 sample determined by the λ-finder method (*w* = 1 mm)

## 4. EXAMPLE II, EFFECTIVE THERMAL CONDUCTIVITY OF PCBS

From the aspect of measurement technology the measure of the effective thermal conductivity of PCBs is greatly simplified by the fact that it can be carried out in the air as well and it is not necessary to use vacuum. In that case, however, the shunting effect of the air must somehow be taken into consideration.

In this experiment we simulated the thermal transient response of 1-mm-thick boards having thermal conductivity between 2 and 14 W/mK in air. Due to the shunting effect of the air the thermal conductivity values of the board simulated in the air determined on the basis of the structure function are, of course, higher than the values adjusted as the parameters of the simulation. The shunting effect of the air changes the deriving thermal conductivity only a little, but the size of this error can be measured in relation to the subjective error appearing at the selection of the two points. In the experiments we employed the following cylinder-symmetrical structure (Figure 7).

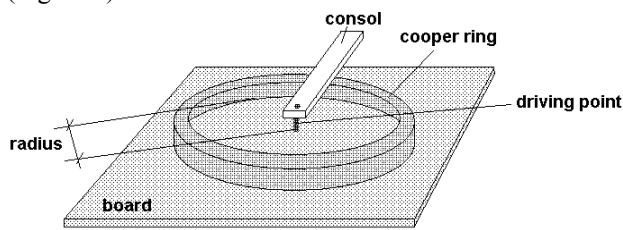

Figure 7: The board thermal conductivity measurement setup

Using the λ-finder algorithm we determined the relative increase of the thermal conductivity values of the board and plotted it in relation to the parameter of the simulation burdened with the shunting effect of the air (Figure 8).

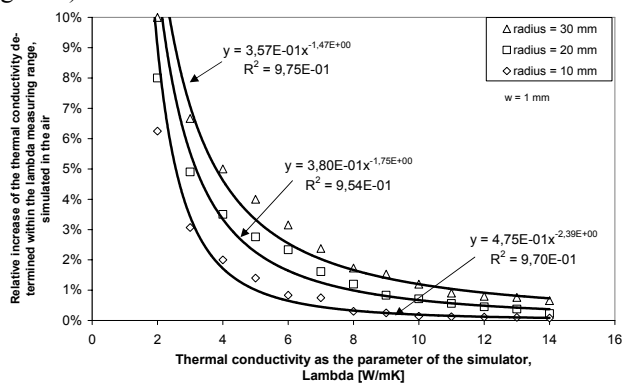

Figure 8: The air surrounding the board increases thermal conductivity

From the aspect of measurement technology it is advisable to use such a group of graphs which gives the thermal conductivity of the sample (that is, measured in vacuum) from the thermal conductivity values measured in the air (Figure 9).

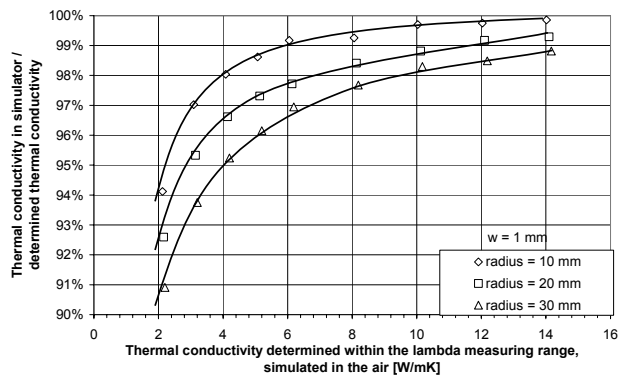

Figure 9: Thermal conductivity adjustment curves of the sample for the easy measuring

## 5. CONCLUSIONS

We have created an algorithm which reduces the subjective error appearing at the determination of the local thermal conductivity based on structure function of board-like materials in a way that it applies a wide range instead of two points of the cumulative structure function to determine the thermal coefficient.

We have shown two applications for the algorithm. This algorithm is advantageous when the subjective error is commensurable to other errors, for example the shunting effect of the air at the determination of the effective thermal conductivity of boards.

## 6. ACKNOWLEDGEMENTS

This work was partially supported by the PATENT IST-1999-12529 Project of the EU, and by the 2/018/NKFP-2001 INFOTERM Projects of the Hungarian Government.